\documentclass{emulateapj}

\lefthead{van der Wel et al.}
\righthead{Axial Ratios of Massive Galaxies}
\slugcomment{The Astrophysical Journal Letters, Accepted}
\begin{document}

\newcommand{\kms}{\>{\rm km}\,{\rm s}^{-1}}
\newcommand{\reff}{R_{\rm{eff}}}
\newcommand{\msol}{M_{\odot}}

\title{Major Merging: The Way to Make a Massive, Passive Galaxy}

\author{Arjen van der Wel\altaffilmark{1}, Hans-Walter
  Rix\altaffilmark{1}, Bradford P.~Holden\altaffilmark{2}, Eric
  F.~Bell\altaffilmark{1,3} \& Aday R.~Robaina\altaffilmark{1}}

\altaffiltext{1}{Max-Planck Institute for Astronomy, K\"onigstuhl 17,
  D-69117, Heidelberg, Germany}

\altaffiltext{2}{University of California Observatories/Lick
  Observatory, University of California, Santa Cruz, California,
  95064, USA}

\altaffiltext{3}{Department of Astronomy, University of Michigan, 500
  Church Street, Ann Arbor, Michigan, 48109, USA}

\begin{abstract}
  We analyze the projected axial ratio distribution, $p(b/a)$, of
  galaxies that were spectroscopically selected from the Sloan Digital
  Sky Survey (DR6) to have low star formation rates.  For these
  quiescent galaxies we find a rather abrupt change in $p(b/a)$ at a
  stellar mass of $\sim10^{11}\msol$: at higher masses there are
  hardly any galaxies with $b/a<0.6$, implying that essentially none
  of them have disk-like intrinsic shapes and must be spheroidal.
  This transition mass is $\sim 3-4$ times higher than the threshold
  mass above which quiescent galaxies dominate in number over
  star-forming galaxies, which suggests that these mass scales are
  unrelated.  At masses lower than $\sim 10^{11}\msol$, quiescent
  galaxies show a large range in axial ratios, implying a mix of
  bulge- and disk-dominated galaxies.  Our result strongly suggests
  that major merging is the most important, and perhaps only relevant,
  evolutionary channel to produce massive ($>10^{11}\msol$), quiescent
  galaxies, as it inevitably results in spheroids.
\end{abstract}


\begin{figure*}[t]
\epsscale{1} 
\plottwo{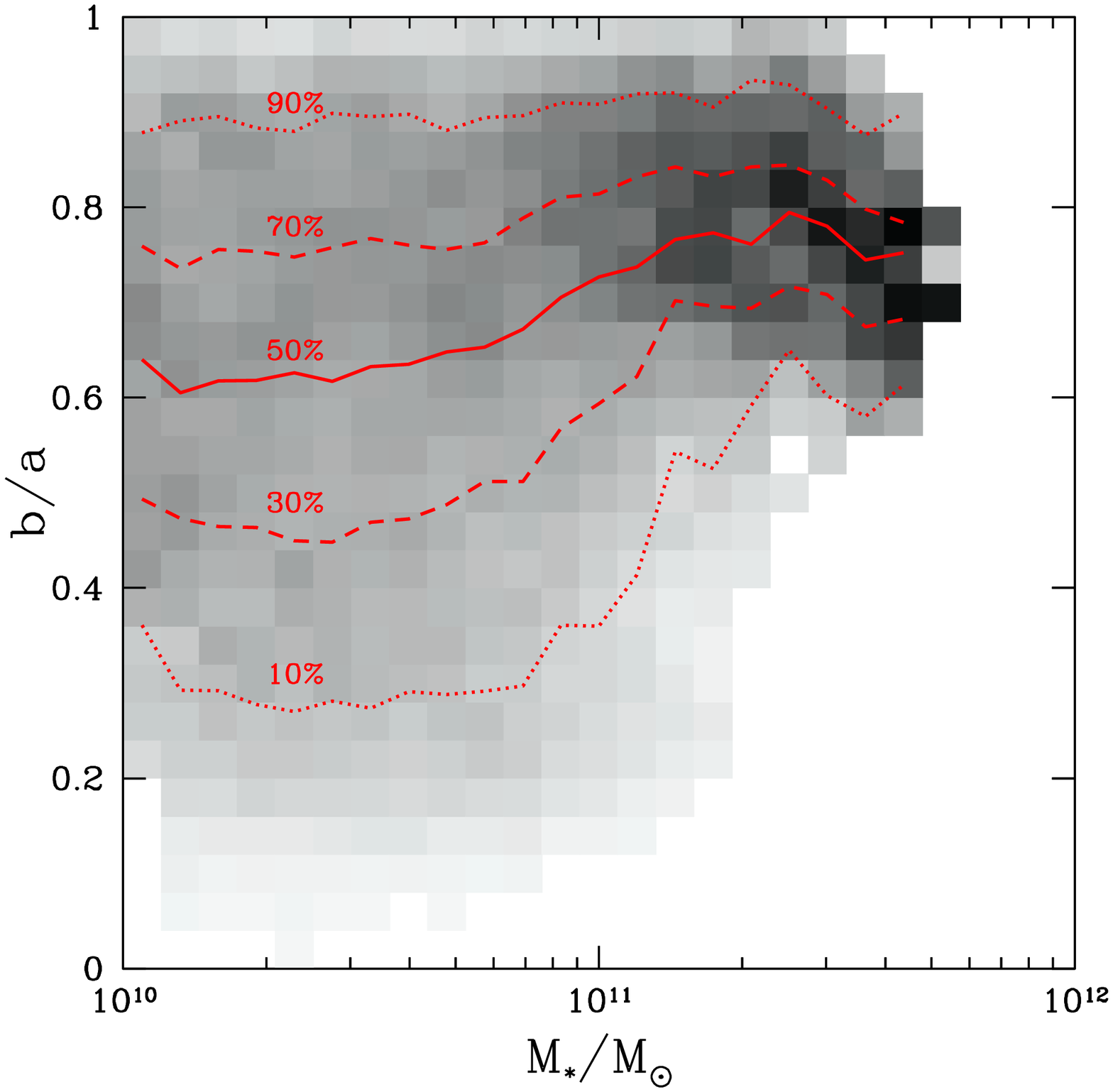}{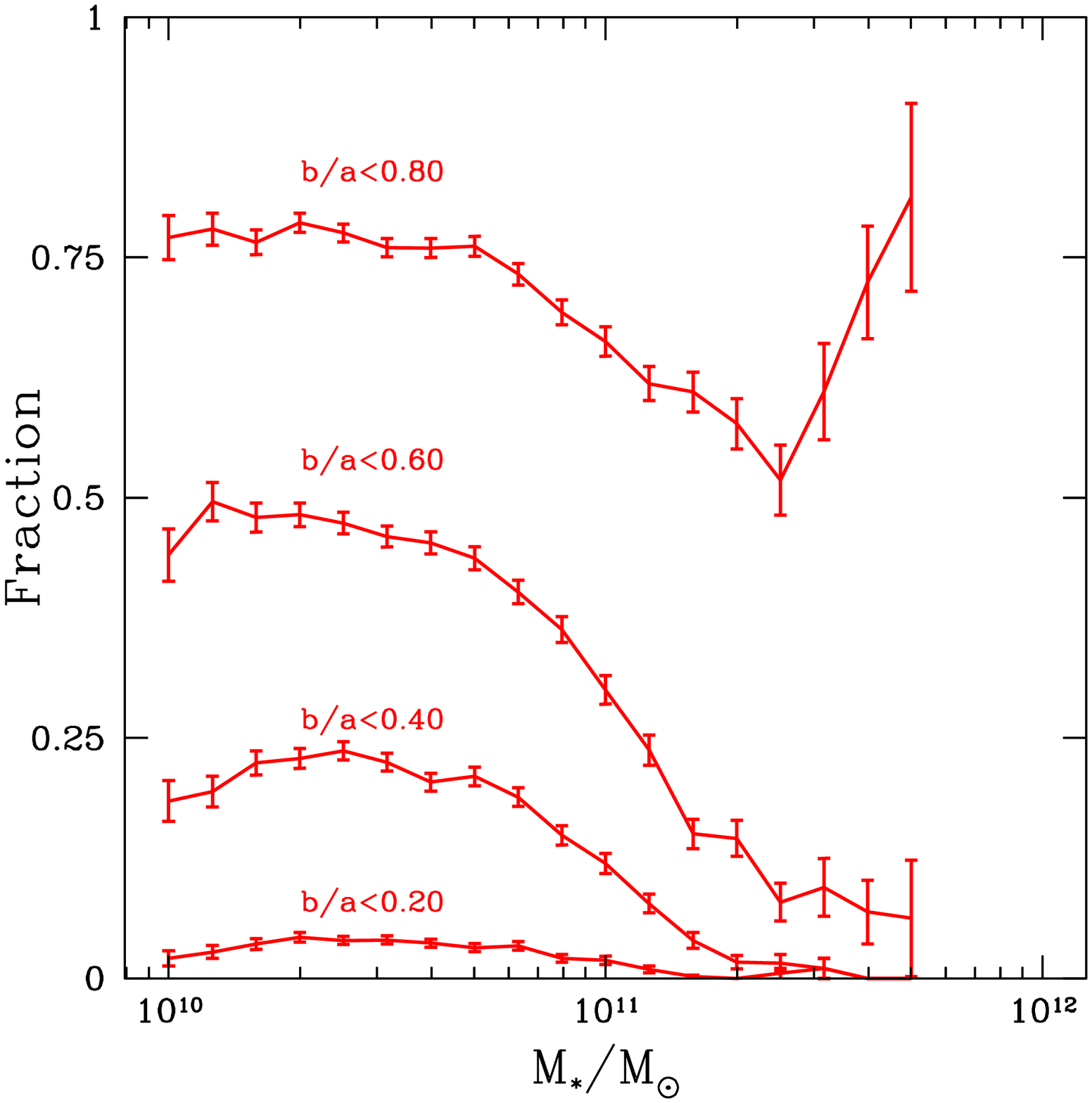}
\caption{\textit{Left:} axial ratio distribution, $p(b/a)$, as a
  function of stellar mass for spectroscopically selected quiescent
  galaxies from the SDSS at $0.04<z<0.08$.  The gray scale represents,
  normalized to the total number of galaxies in narrow bins of stellar
  mass, the fraction of galaxies with axial ratio $b/a$. The upper
  boundaries below which, as a function of mass, 10\%, 30\%, 50\%,
  70\%, and 90\% of galaxies are located, are delineated by the red
  lines.  \textit{Right:} fraction of galaxies with axial ratios $b/a$
  smaller than 0.8, 0.6, 0.4, and 0.2, as a function of stellar mass.
  These figures clearly show that at $M_*\ge 10^{11}\msol$, the
  fraction of galaxies with small $b/a$ decreases rapidly with mass.
  At lower masses, $p(b/a)$ is approximately uniform in the range
  $0.3<b/a<0.9$, implying a significant contribution of disks.  At
  higher masses, axial ratios are approximately evenly distributed in
  the range $0.6<b/a<0.9$, which shows that disks must be rare, and
  galaxies intrinsically round.}
\label{M_q}
\end{figure*}

\section{Introduction}
Even galaxies with little star formation activity continue to evolve,
as evidenced by the substantial increase of their cosmic stellar mass
density over the past 7 billion years \citep{bell04b, faber07,
  brown07}.  This must be related to the decreasing star formation
activity over the same period \citep[e.g.,][]{lefloch05}, and the
production of such quiescent galaxies through the truncation of star
formation \citep[e.g.,][]{faber07, bell07}; the color scatter among
quiescent galaxies and its evolution are in precise agreement with
such a scenario \citep{ruhland09}.  There are, however, quiescent
galaxies at all redshifts $z\lesssim 1.3$ that are more massive than
the most massive star-forming galaxies. This implies that star
formation in the most massive galaxies was truncated even earlier,
and/or that mergers play an important role in producing massive
galaxies.

Evidence for the early formation of massive galaxies is provided by
their old stellar populations. However, we need to bear in mind that
there can be a large difference between the age of the stellar
population and the assembly age, especially if mergers are important,
as is the case in a hierarchical framework for galaxy formation
\citep{delucia07}.  Hence, the number density evolution of galaxies is
important in constraining their assembly history.  Measuring this is
difficult because of its sensitivity to the luminosity evolution
correction, especially for massive galaxies at the exponential cut-off
of the mass function.  As a result, there is no consensus among the
currently available measurements \citep{cimatti06, wake06, brown07,
  cool08}.

Given these difficulties, other observations have been used to either
directly or indirectly constrain the assembly of galaxies.  Merging
activity among the massive galaxy population is observed
\citep[e.g.,][]{vandokkum99, vandokkum05, bell06a, bell06b, lin08},
and has been shown to produce a color-magnitude relation that is in
agreement with observations \citep{skelton09}.  However, its
cosmological relevance has always been difficult to determine, given
the uncertainties in converting observed merger fractions to merger
rates and the associated growth in mass.  An independent and indirect
indication that massive galaxies undergo continuous evolution is
provided by the recent result that high-redshift quiescent galaxies
are substantially smaller than local galaxies with the same mass
\citep[see,][and references therein]{vanderwel08c}.  This strongly
suggests that mergers are important \citep[see,
e.g.,][]{vanderwel09a}, and that the assembly of massive galaxies is
continuing up until the present day.  Another indirect, yet powerful,
constraint is provided by the evolution in the clustering and halo
occupation distribution of red galaxies \citep{white07, conroy07,
  brown08}: the evolution in the clustering strength of red galaxies
is slower than expected in the absence of merging.

In this Letter we address the question whether major merging is the
dominant mechanism for the production of very massive, quiescent
galaxies.  The argument that we invoke is simply that major merging
generally leads to rounder galaxies.  An analysis of the shape
distribution of quiescent galaxies can therefore constrain the
importance of merging.  Since merging among galaxies with mass ratios
of $\lesssim 3$ is the only known mechanism to produce round galaxies
(see Section 3 for further discussion), this is a powerful test.  The
disadvantage of this method, compared to those mentioned above, is
that no information about the time scale and epoch of galaxy assembly
can be inferred.

\citet{vincent05} and \citet{padilla08} were the first to
systematically study the axial ratio distribution, $p(b/a)$, of a
large number of galaxies, selected from the Sloan Digital Sky Survey
(SDSS).  Through a detailed analysis, they infer the intrinsic shape
distribution and the effect of extinction.  Both divide the sample
into 'elliptical' and 'spiral' galaxies, and confirmed that luminous
'elliptical' galaxies are, on average, rounder and tri-axial, compared
to low-luminosity 'ellipticals', which are more elongated and oblate
\citep{davies83, franx91}, and display disky isophotes
\citep{jorgensen94}. This phenomenon is not recent: \citet{holden09a}
showed that this trend persists at least out to $z\sim 1$.  Here we
present a complementary, modified analysis, focusing on $p(b/a)$ as a
function of stellar mass for quiescent, i.e., non-star-forming,
galaxies.  Because mass-to-light ratios are well constrained by
broad-band colors for quiescent galaxies, stellar mass estimates are
robust.  This is essential for our purposes, as we are interested in
the most massive objects, i.e., those that populate the exponential
tail of the mass function.  Furthermore, as opposed to previous
studies, we pre-select galaxies independent of their photometric
properties.  Our shape-independent, spectroscopic selection criteria
circumvent the biases that are potentially introduced by selecting
galaxies by their 'morphological' properties, or some pre-defined
surface brightness profile.

With this sample, for which we have determined axial ratios from our
own fits to two-dimensional light distributions, we address the
following specific questions. Are high-mass, quiescent galaxies
rounder than low-mass quiescent galaxies? If so, is there a mass limit
at which $p(b/a)$ distinctly changes, and above which disk-dominated
are completely absent?  Such evidence would imply that the only
evolutionary path to such masses is a disk-destroying mechanism, i.e.,
major merging.

\section{The Sample}

We select a sample of 17,480 quiescent galaxies from Data Release 6 of
the SDSS \citep{adelman08}.  Our sample includes galaxies at redshifts
$0.04<z<0.08$ without detectable $[\rm{OII}]$ and $\rm{H}\alpha$
emission lines. The selection criteria are described and motivated in
full by \citet{graves09b}; but as opposed to that work, we do not
exclude galaxies with a low concentration index and galaxies that are
fit better by an exponential profile than by a \citet{devaucouleurs48}
profile, because this may exclude quiescent, yet disk-like galaxies,
which are obviously relevant for quantifying $p(b/a)$ of quiescent
galaxies. As a consequence, our sample may include galaxies with star
formation in an extended disk outside the SDSS spectroscopic fiber.
This effect, however, does not affect our main conclusion that
quiescent massive galaxies with prominent disks are extremely rare
(see Section \ref{res}). Rather, such a bias works in the opposite
direction in the sense that it would lead to the mistaken inclusion of
galaxies with large disks.

The exclusion of all galaxies with emission lines also excludes
quiescent galaxies with active galactic nuclei.  Their number,
however, is small, and make up a small fraction of the population
\citep[e.g.,][]{pasquali09a} that is negligible for our purposes.

The axial ratios were obtained as described by \citet{vanderwel08c}.
Briefly, GALFIT \citep{peng02} is used to determine from the $r$-band
the radii, axial ratios, position angles, and total magnitudes,
assuming a \citet{devaucouleurs48} surface brightness profile.  We
have verified that adopting surface brightness models with a free
S\'ersic index does not lead to a significantly different
$p(b/a)$.

The stellar masses are derived with the simple conversion from color
to mass-to-light ratio \citep{bell03}, but are normalized to
correspond to the \citet{kroupa01} stellar initial mass function.  The
assumed cosmology is $(\Omega_{\rm{M}},~\Omega_{\Lambda},~h) =
(0.3,~0.7,~0.7)$.

The sample is complete over the entire redshift range $0.04 < z <
0.08$ down to $M_* \sim 4\times 10^{10}\msol$, set by the
spectroscopic magnitude limit of the SDSS ($r=17.7$).  The SDSS may be
incomplete for low-luminosity, low-surface brightness galaxies
\citep{blanton05c}, which could, in addition, depend on their
orientation \citep[see, e.g.,][]{odewahn97}.  However, since we are
concerned with the high-mass end of the galaxy population, this does
not play a role.  Moreover, simulations of images with even lower
signal-to-ratio than those of the massive galaxies analyzed here
demonstrate that axial ratio measurements from GALFIT are robust and
accurate \citep{holden09a}.  In summary, the lack of galaxies with
small $b/a$, reported in the following section, is not in any way
compromised by selection effects or measurement errors.

\section{Results and Discussion}\label{res}

In Figure \ref{M_q}(a) we show $p(b/a)$ of the 17,480
spectroscopically selected, quiescent galaxies as a function of
stellar mass.  $p(b/a)$ is shown in gray scale, with the percentiles
of the cumulative $b/a$ distribution shown as (red) lines.  Figure
\ref{M_q}(a) immediately demonstrates that for quiescent galaxies, the
projected axial ratio distribution is a strong function of stellar
mass. In the narrow mass range $8\times 10^{10} \lesssim M_*/M_{\odot}
\lesssim 2\times 10^{11}$ there is a rapid decrease in the number of
galaxies with small axial ratios.  As further illustrated by Figure
\ref{M_q}(b), above $M_*\sim 2\times 10^{11}~M_{\odot}$ quiescent
galaxies with $b/a<0.6$ are essentially absent.

This result shows that evolutionary paths that lead to quiescent
galaxies with stellar mass $M_* \gtrsim 2\times 10^{11}~M_{\odot}$ all
but exclude the existence, or the survival, of highly flattened,
disk-like stellar components.  As highly flattened stellar systems are
quite common at lower masses, in the possible realm of plausible
progenitors of high-mass galaxies, this result implies the destruction
of the flattened component in whatever process causes growth beyond
$M_*\sim 2\times 10^{11}~\msol$.  Therefore, our result that
essentially all quiescent galaxies with masses larger than $M_*\sim
2\times 10^{11}~\msol$ are round strongly suggests that for such
galaxies major mergers are the dominant, perhaps even unique,
formation channel.  The destruction of a stellar disk requires a major
merger, i.e., a merger involving progenitors with a relatively small
mass ratio of at most $\sim 3$, mergers with a larger mass ratio
leaving stellar disks intact \citep[see, e.g.,][]{bekki98,
  bournaud04}.  Moreover, most likely, the progenitors are not very
gas rich, as this would produce a disky remnant
\citep[e.g.,][]{naab06b}.

It has been suggested that cold flows are responsible for the
formation of massive, classical bulges at high redshift
\citep{dekel09, ceverino09}.  In this scenario, intensely star-forming
'knots' merge, forming a massive bulge \citep[see also][]{noguchi99}.
However, a substantial fraction of the mass ($\sim 50\%$) is still
predicted to reside in a disk.  Even at later stages, when the gas
disk has become stable against fragmentation and collapse
\citep[``morphological quenching'';][]{martig09}, the stellar disk
remains intact and contains a non-negligible fraction of the total
mass.  In short, although cold flows plausibly produce quiescent
galaxies, the end-products will not be uniquely round.  Only in the
case of sufficient merger activity would galaxies become spheroidal.

In passing, we note that the sharp decrease in the fraction of very
round galaxies ($b/a > 0.8$) at the very highest masses ($M>3\times
10^{11}~\msol$) signifies that such high mass galaxies are typically
brightest group/cluster galaxies, which tend to be slightly more
elongated than 'normal' massive elliptical galaxies
\citep[see,][]{bernardi08}.

As already noted above, at masses lower than $M_*\sim 10^{11}~\msol$,
quiescent galaxies display a large range in axial ratios, which
implies that star formation truncation mechanisms below
$10^{11}~\msol$ are often not associated with the destruction of the
disk.  It remains to be tested whether $p(b/a)$ of low-mass quiescent
galaxies is similar to or different from $p(b/a)$ of star-forming
galaxies in the same mass range. Such an analysis, which is
non-trivial because of the effects of extinction and color gradients,
will constrain the degree to which mergers or bulge growth regulate
star formation at these lower masses.

Another open question concerns the number and properties of massive,
star-forming galaxies.  Morphological studies suggest that a large
fraction ($20\%-40\%$) of all galaxies more massive than $M\sim
10^{11}~M_{\odot}$ are late-type galaxies \citep{vanderwel08a,
  bamford09}, and their high masses of at least some of these objects
are confirmed by their rotational velocities
\citep[e.g.,][]{courteau07}. Yet, the degree to which such galaxies
are disk-dominated and should be considered actively star forming
remains to be determined.  It would, therefore, be premature to
conclude that merging is the only way to produce a massive galaxy in
general, and therefore restrict this proposition to the formation of
massive, quiescent galaxies.

The picture sketched by the axial ratio distribution of quiescent
galaxies is in agreement with the strong correlation between structure
and mass for galaxies in general \citep[e.g.,][]{kauffmann03b,
  vanderwel08a} and early-type galaxies in particular
\citep[e.g.,][]{caon93, graham03}: high-mass galaxies are more
concentrated and have higher S\'ersic indices than low-mass galaxies.
These trends are an indirect indication of a decreasing importance of
disks for galaxies with higher masses, although part of this trend is
caused by the increase in S\'ersic index with galaxy mass among
spheroidal galaxies.  In our sample we see a similar trend: in the
mass range $10^{10} < M_*/\msol < 2\times 10^{10}$, 41\% of the
galaxies have S\'ersic indices $n<3$, whereas at higher masses,
$M_*>2\times 10^{11}\msol$, only 3\% have such low S\'ersic
indices. We postpone a full exploration of the joint behavior of shape
and structure as a function of galaxy mass until a future paper, but
it is encouraging that the apparent absence of prominent disks in
high-mass, quiescent galaxies is reflected in both the S\'ersic index
and $p(b/a)$.

\acknowledgements{The authors thank the referee for positive feedback.
  A.v.d.W. thanks Marijn Franx for helpful discussion.  H.W.R. thanks
  Simon White for his insistence on eventually 'publishing the last
  thesis chapter'. This Letter makes use of the Sloan Digitial Sky
  Survey (www.sdss.org).}

\bibliographystyle{apj}

\end{document}